\documentclass[aps,prx,reprint,superscriptaddress,amsmath,amssymb]{revtex4-2}
\usepackage{color}
\usepackage{amsmath,amssymb}
\usepackage{graphicx}
\usepackage{ulem}
\usepackage[colorlinks, linkcolor=blue, citecolor=blue, urlcolor=blue, breaklinks=true]{hyperref}
\usepackage{subfig}
\usepackage{mathtools}

\DeclarePairedDelimiter\bra{\langle}{\rvert}
\DeclarePairedDelimiter\ket{\lvert}{\rangle}
\DeclarePairedDelimiterX\braket[2]{\langle}{\rangle}{#1\,\delimsize\vert\,\mathopen{}#2}

\begin{document}
\title{How gravitational waves change photon orbital angular momentum quantum states}
\author{Haorong Wu}
\affiliation{School of Physics and Technology, Wuhan University, Wuhan 430072, China}
\author{Xilong Fan}
\email{xilong.fan@whu.edu.cn}
\affiliation{School of Physics and Technology, Wuhan University, Wuhan 430072, China}
\author{Lixiang Chen}
\affiliation{Department of Physics, Xiamen University, Xiamen 361005, China}

\begin{abstract}
We explore the evolution of vortex light in the presence of gravitational waves (GWs) and demonstrate that the quantized orbital angular momentum (OAM) states can make transitions to other states due to the GWs. The interaction is calculated based on the framework of the wave propagation in linearized gravity theory and canonical quantization of the light field in curved spacetime. It is found that when a photon possessing OAM of $l$ interacts with GWs, the OAM modes of $l\pm1$ and $l\pm2$ may be excited with probabilities of $P_{l\pm1}\sim 10^{-17}$ and $P_{l\pm2}\sim 10^{-20}$, respectively. Higher probabilities of the transitions can be achieved when the photon radial wave vector or the propagation distance is increased, or when the photons encounter GWs with stronger amplitudes or smaller frequencies.  Thus, a new GW detection technique is proposed, which may exhibit good performance in a wide range of GW frequencies. Furthermore, the detector is insensitive to seismic noise and is more advantageous for determining the distance of the source compared to current interferometer detectors.
\end{abstract}
\maketitle

\section{Introduction}
Gravitational waves (GWs) are ripples in the fabric of spacetime, generated by accelerating objects, such as mergers of binary black holes, coalescences of binary neutron stars, spinning neutron stars, or supernova explosions, among others \cite{flanagan2005basics}. They may also have been produced during the early stage of the universe, serving as a residue of the big bang as the cosmic microwave background \cite{caprini2018cosmological}. The majority of documented GW events are detected using interferometric instruments. These detectors can be categorized into three generations according to the technology employed \cite{punturo2010third}. On the other hand, the majority of GW detection efforts are concentrated in four frequency bands \cite{ni2018gravitational}: the high-frequency band (10 Hz to 100 kHz) by ground-based laser-interferometric detectors \cite{abbott2016observation}, the upper part (10 $\mu$Hz to 0.1 Hz) of the low-frequency band by space laser-interferometric detectors \cite{amaro2012low}, the very low-frequency band (nHz band, 300 pHz to 100 nHz) by pulsar timing arrays \cite{hobbs2010international}, and the extremely low-frequency band (cosmological band; 1 aHz to 10 fHz) by cosmic microwave background experiments \cite{wu2014guide}. However, the detection of GWs in the mid-frequency band (0.1 Hz to 10 Hz) remains elusive \cite{ni2018gravitational}.

Moreover, contemporary interferometric detectors have not completely utilized all sources available from the laser, particularly the orbital angular momentum (OAM) degrees of freedom. The OAM of light is related to the field spatial distribution. Since Allen {\it et al.} \cite{allen1992orbital} first proposed the OAM of light in 1992, it has received a significant amount of attention, and the findings of investigations that followed have ushered in a new age of modern optics \cite{allen1992orbital,calvo2006quantum,mair2001entanglement,molina2001management,leach2002measuring,vaziri2002experimental,vaziri2003concentration,allen1999iv,kivshar2001optical,molina2007twisted,yao2011orbital,molina2001propagation}. In recent years, there were several attempts to connect the photon OAM detection with astronomical observations, such as possible astronomical apparatus for photon OAM and their limitations \cite{harwit2003photon}, direct observational demonstration of the existence of rotating black holes \cite{tamburini2011twisting,tamburini2020measurement}, and effects of reduced spatial coherence of astronomical light sources on the OAM detection \cite{hetharia2014spatial}. Meanwhile, there has been a growing interest in the intersection of GW and OAM fields. To name a few, some researchers have studied the possibilities of GWs with OAM \cite{bialynicki2016gravitational,baral2020gravitational,bialynicki2018trapping}. It is proposed and demonstrated that higher-order Laguerre-Gauss modes can be used in GW detectors to significantly reduce the thermal noise \cite{PhysRevD.79.122002,PhysRevLett.105.231102}. The helicity coupling of structured lights to GWs is studied, and it is found that the coupling effect between angular momenta of the GWs and twisted lights may make photons undergo transitions between different OAM eigenstates and lead to some measurable optical features by Wu {\it et al.} \cite{wu2022testing}. However, their work is limited by the paraxial approximation and the assumption that the GW is a static gravitational field, discarding its waveform. The question as to how the GW background affects the coherence and the degree of high-dimensional OAM entanglement when twisted photons travel across the textures of curved spacetime is studied by Wu {\it et al.} \cite{PhysRevD.106.045023}.

The prevalence of GWs, the significance of GW detection, and the numerous sources afforded by the quantum characteristics of OAM prompt a fundamental question as to the propagation of vortex beams in GWs and the potential enhancement of GW detection through the quantum nature of OAM. We employ the 3+1 dimensional Green's function method in cylindrical coordinates to analyze the propagation of vortex beams in GWs, deriving the perturbation light field. The evolution of quantum OAM states in GWs is examined through the Bogoliubov transformation, revealing the interaction between GWs and vortex beams. Based on the result, we propose a photonic single-arm GW detector by utilizing the quantum characteristics of OAM photons. Unless otherwise specified, units with $c=G=\hbar=1$ are used. The metric signature is chosen as $(-,+,+,+)$. All Greek indices run in $\{0, 1, 2, 3\}$. We will use $\delta(x)$ to represent a Dirac delta function when $x$ is a continuous variable, and to represent a Kronecker delta function when $x$ is discrete.

\section{3+1 dimensional Green's function in cylindrical coordinates}
It is known that the wave equation,\begin{equation}
	(-\partial^2_0+\partial^i\partial_i)\psi(x^\mu)=f(x^\mu),\label{eq.waveEquation}
\end{equation}
can be solved by \cite{duffy2015green}\begin{equation}
	\psi(x^\mu)=\int d^4 x^{\mu'}\mathcal G(x^\mu,x^{\mu'}) f(x^{\mu'}),\label{eq.greenFun}
\end{equation}
where the Green's function $\mathcal G(x^\mu,x^{\mu'})$ satisfies \begin{align}
&	(-\partial^2_0+\partial^i\partial_i)\mathcal G(x^\mu,x^{\mu'})=\delta(x^\mu-x^{\mu'})\nonumber \\=&\delta(t-t') \rho^{-1} \delta(\rho-\rho')\delta(\phi-\phi')\delta(z-z'),
\end{align}
in cylindrical coordinates. Using the fact that
\begin{equation}
\frac 1 \rho\delta(\rho-\rho')=\int_0^\infty dk_\perp k_\perp J_m(k_\perp \rho)J_m(k_\perp \rho')	,
\end{equation}
with $J_m(x)$ being the $m$th-order Bessel function \cite{lozier2003nist,olver2010nist},
\begin{equation}
	\delta(\phi-\phi')=\frac 1{2\pi}\sum_{m=-\infty}^\infty \exp [im(\phi-\phi')],
\end{equation}
and \begin{equation}
	\delta(z-z')=\frac 1{2\pi}\int_{-\infty}^{\infty}dk_3 \exp[ik_3(z-z')],
\end{equation}
we can write\begin{align}
	\mathcal G(x^\mu,x^{\mu'})=&\frac{1}{4\pi^2}\sum_{m=-\infty}^\infty \int_{-\infty}^\infty dk_3 \int_0^\infty d k_\perp e^{ik_3(z-z')} \nonumber \\ &\times e^{im(\phi-\phi')} k_\perp J_m(k_\perp \rho) J_m(k_\perp \rho')g_\omega(t,t'), \label{eq.defG}
\end{align}
where $g_\omega(t,t')$ satisfies $(-\partial^0_0-\omega^2)g_\omega(t,t')=\delta(t-t')$, $\omega=(k^2_\perp+k^2_3)^{1/2}$, $k_3$ is the $z$ component of wavevector, and $k_\perp$ is the radial wavevector.

To find the expression of $g_\omega(t,t')$, let us first suppose the source $f(x^\mu)$ can be separated into a spatial part $f_x(\mathbf x)$ and a temporal part $f_t(t)$. Hence, $f(x^\mu)=f_x(\mathbf x)f_t(t)$. This separation of $f(x^\mu)$ is suitable in our study, since the scalar fields and GWs can be written in this separation form. From this separation and Eqs. \eqref{eq.greenFun} and \eqref{eq.defG}, one can deduce that the integration in Eq. \eqref{eq.greenFun} can be separated into a spatial part and a temporal part, as well. The spatial integration will give the spatial part of $\psi(x^\mu)$, denoted by $\psi_x(\mathbf x)$. Meanwhile, the temporal integration will give the temporal part as $\psi_t(t)=\int dt' g_\omega(t,t') f_t(t')$. These two parts constitute $\psi(x^\mu)$---i.e., $\psi(x^\mu)=\psi_x(\mathbf x)\psi_t(t)$. The function $\psi_t(t)$ satisfies $(-\partial^2_0-\omega^2)\psi_t(t)=f(t)$. The corresponding homogeneous equation $(-\partial^2_0-\omega^2)\tilde \psi_t(t)=0$ has two linearly independent solutions $\tilde \psi_1(t)=\exp (i\omega t)$ and $\tilde \psi_2(t)=\exp (-i\omega t)$. Then, $g_\omega(t,t')$ can be written as a linearly superposition of $\tilde \psi_1(t)$ and $\tilde \psi_2(t)$---i.e.,\begin{equation}
	g_\omega(t,t')=c_1(t') \tilde \psi_1(t)+c_2(t') \tilde \psi_2(t).
\end{equation}
To identify $g_\omega(t,t')$, we need to impose some boundary conditions for $\psi_t(t)$. Different conditions will give different $g_\omega(t,t')$. Here, we choose to use the homogeneous boundary condition at $t=0$---i.e., $\psi_t(0)=0 $ since we will suppose the GW starts to interact with the scalar field at $t=0$, in the following section. For the case where $t<t'$, we can choose $g_\omega(t,t')=0$, so that the homogeneous boundary is satisfied as \begin{align}
	\psi_t(0)=&\left .\int_0^\infty dt' g_\omega(t,t') f(t') \right |_{t=0}\nonumber \\=&\left .\left [ \int_0^t dt' g_\omega(t,t') f(t')+\int_t^\infty dt' g_\omega(t,t') f(t') \right ] \right |_{t=0}\nonumber \\=&0.
\end{align}
For the other case where $t>t'$, there are two undetermined coefficients, $c_1(t')$ and $c_2(t')$, in $ g_\omega(t,t')$, so two linearly independent equations are needed to identify $g_\omega(t,t')$. The first equation is given by the continuity of $g_\omega(t,t')$ at $t=t'$---i.e., near the point $t=t'$, we have $g_\omega(t'-\epsilon,t')=g_\omega(t'+\epsilon,t')$, with $\epsilon$ being an infinitesimal positive number only in this section. Hence, one can have
\begin{equation}
	c_1(t') e^{i\omega t'}+c_2(t') e^{-i\omega t'}=0,\label{eq.g1}
\end{equation}
where $g_\omega(t'-\epsilon,t')=0$ is used. The second equation can be derived by noting that\begin{equation}
	\int_{t'-\epsilon}^{t'+\epsilon}dt(-\partial^2_0-\omega^2)g_\omega(t,t')=\int_{t'-\epsilon}^{t'+\epsilon}dt\delta(t-t')=1,
\end{equation} 
so one can have\begin{align}
	&\int_{t'-\epsilon}^{t'+\epsilon}dt(-\partial^2_0)g_\omega(t,t')\nonumber \\=&-\left [\left .\frac {\partial g_\omega(t,t')}{\partial t} \right |_{t'+\epsilon}-\left .\frac {\partial g_\omega(t,t')}{\partial t} \right |_{t'-\epsilon}\right ] \nonumber \\=&-i\omega c_1(t') e^{i\omega t'}+i\omega c_2(t')e^{-i\omega t'}=1,\label{eq.g2}
\end{align}
where $\int_{t'-\epsilon}^{t'+\epsilon}dt \omega^2 g_\omega(t,t')=0$ is used. By combining Eqs. \eqref{eq.g1} and \eqref{eq.g2}, $c_1(t')$ and $c_2(t')$ can be solved to be
\begin{align}
	c_1(t')=&\frac i{2\omega}e^{-i\omega t'},\\
	c_2(t')=&\frac {-i}{2\omega}e^{i\omega t'}.
\end{align}
Therefore, for $t>t'$, one can write\begin{equation}
	g_\omega(t,t')=\frac 1{2i\omega}(\exp [-i\omega(t-t')]-\exp [i\omega(t-t')]).
\end{equation}
In total, the function $g_\omega(t,t')$ can be expressed as\begin{equation}
	g_\omega(t,t')=\frac {\theta(t-t')} {2i\omega}\left (\exp [-i\omega(t-t')]-\exp [i\omega(t-t')]\right ),
\end{equation}
where $\theta(t-t')$ is the Heaviside step function. This equation can be further written as a contour integration \cite{peskin2018introduction}. 

\begin{figure} [tbhp]
	\centering
	\includegraphics[width=0.8\linewidth]{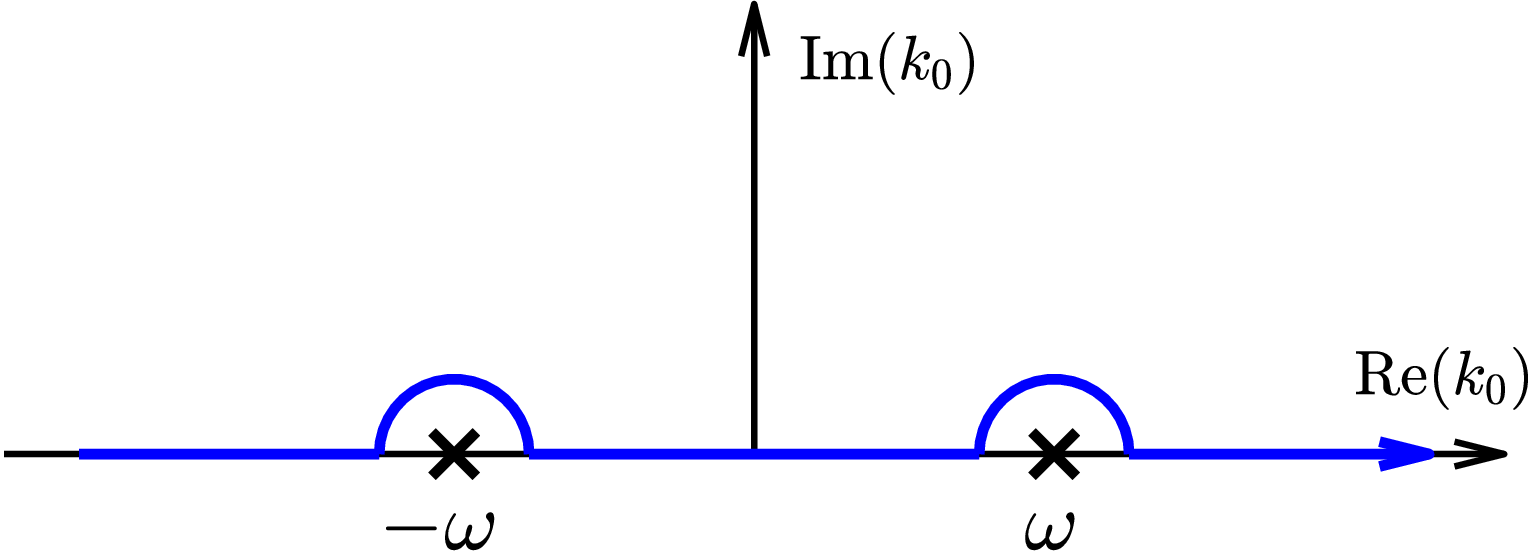}
	\caption{The path for integral with respect to $k_0$ in Eq. \eqref{eq.GreenFunctionIntegral}. The contour is closed in the lower half plane.}
	\label{fig.ContourIntegral}
\end{figure}

Therefore, the 3+1 dimensional Green's function in cylindrical coordinates is given by\begin{align}
	\mathcal G(x^\mu,x^{\mu'})=&\frac{1}{8\pi^3}\sum_{m=-\infty}^\infty \int_{-\infty}^\infty dk_3 \int_0^\infty d k_\perp\int dk_0 \frac{k_\perp }{k^2_0-\omega^2}\nonumber \\ &\times  e^{-ik_0(t-t')+ik_3(z-z')+im(\phi-\phi')} J_m(k_\perp \rho) \nonumber \\ &\times J_m(k_\perp \rho')\theta(t-t'),\label{eq.GreenFunctionIntegral}
\end{align} 
where the integral with respect to $k_0$ is a contour integration, along the path shown in Fig. \ref{fig.ContourIntegral}.

\section{Perturbation of vortex beams in GWs}
It has been demonstrated that under the Lorentz gauge, each component of photons in curved spacetime evolves as if it were a massless scalar field \cite{exirifard2021towards,chakraborty2024particle,morales2020scalar}. Consequently, to maintain simplicity and concentrate on the orbital angular momentum (OAM) degrees of freedom, we opt to employ a massless scalar field to represent the vortex beam, which propagates along the positive $z$ axis. For GWs propagating along $z$ axis, it can be written as, in the TT gauge,
\begin{equation}
	h^{\rm TT}_{\alpha\beta}(x^\mu)=A_\lambda \cos [k_g(-t+z)+\phi_0]\tilde e^\lambda_{\alpha\beta},
\end{equation}
where $\lambda=(+,\times)$ is the GW polarization, $A_\lambda$ represent amplitudes for the two polarizations, $k_g$ is the GW frequency, $\phi_0$ is the initial phase, and $\tilde e^\lambda_{\alpha\beta}$ are the unit linear-polarization tensors for GWs \cite{misner1973gravitation}. We absorb the amplitudes into the polarization tensor as \begin{equation}
	\tilde \epsilon_{\alpha\beta}=A_\lambda \tilde e^\lambda_{\alpha\beta}=\begin{pmatrix}
		0&0&0&0\\
		0&A_+&A_\times&0\\
		0&A_\times&-A_+&0\\
		0&0&0&0
	\end{pmatrix}.
\end{equation}
For a general GW, given that the vortex beam exhibits axial symmetry around its propagation axis \cite{andrews2021symmetry}, we can, without loss of generality, rotate the coordinates to allow the GW to propagate at an angle $\theta$ with respect to the $z$ axis, situated within the $x-z$ plane. We might further adjust the coordinates to ensure that the GW begins to interact with the vortex beam at $t=0$. Hence, in the linearized gravity theory, the spacetime metric is given by $g_{\alpha\beta}(x^\mu)=\eta_{\alpha\beta}+\epsilon h^{\rm TT}_{\alpha\beta}(x^\mu)$, where $\eta_{\alpha\beta}$ is the Minkowski metric, and $\epsilon$ is a systematic perturbation symbol used to keep track of the number of times the perturbation $h^{\rm TT}_{\alpha\beta}(x^\mu)$ enters. The metric perturbation $h^{\rm TT}_{\alpha\beta}(x^\mu)$ is a very small quantity, as its amplitude can be smaller than $10^{-20}$ when the GWs pass through the earth. When $\epsilon=0$, the result of this study will reduce to the case where no GWs exist; if $\epsilon=0.5$, the result corresponds to the interaction between light and GWs with half amplitudes. Last, $\epsilon=1$ means that the result is calculated with the amplitudes of GWs unaltered. Therefore, at the end of the calculation, we could set $\epsilon\rightarrow 1$ to get back to the full-strength problem of $h^{\rm TT}_{\alpha\beta}(x^\mu)$ \cite{sakurai2020modern,exirifard2021towards}. The perturbation metric field $h^{\rm TT}_{\alpha\beta}(x^\mu)$ is now given by
\begin{align}
	h^{\rm TT}_{\alpha\beta}(x^\mu)=&\frac {\epsilon_{\alpha\beta}}2 [e^{ik_g(-t+x\sin\theta +z\cos\theta )+i\phi_0}\nonumber \\ &+e^{-ik_g(-t+x\sin\theta +z\cos\theta )-i\phi_0}],
\end{align} 
where $\epsilon_{\alpha\beta}$, in cylindrical coordinates $(t,\rho,\phi,z)$, is related to $ \tilde \epsilon_{\alpha\beta}$ by\begin{equation}
	\epsilon^{\mu\nu}=\Lambda^{\mu}_{~\alpha} \Lambda^{\nu}_{~\beta} R^{\sigma}_{~\alpha} R^{\gamma}_{~\beta} \tilde \epsilon^{\sigma\gamma},
\end{equation} 
where $R^{\mu}_{~\alpha}$ are the rotation matrices, and $\Lambda^{\mu}_{~\alpha}$ are the coordinate transformation matrices (from Cartesian to cylindrical coordinates). The cosine function in $h^{\rm TT}_{\alpha\beta}(x^\mu)$ has been split into two exponential functions. In the following computation, we will first deal with \begin{equation}
	h_{\alpha\beta}(x^\mu)=\epsilon_{\alpha\beta} \exp[ik_g(-t+x\sin\theta +z\cos\theta )+i\phi_0].
\end{equation}
After the result is derived, we will repeat the calculation with $k_g$ and $\phi_0$ replaced by $-k_g$ and $-\phi_0$, respectively. The final result will be the average of the two cases.

The action of massless scalar fields in GWs is given by \begin{equation}
	S[\psi(x^\mu)]=\int d^4x^\mu (-g)^{1/2}\frac 1 2g^{\alpha\beta} \psi(x^\mu)_{;\alpha}\psi(x^\mu)_{;\beta},
\end{equation}
with $g=\det(g_{\mu\nu})$. In this action, the scalar fields are minimally coupled to the curved spacetime---i.e., the direct interaction with the spacetime curvature scalar $R$ is turned off \cite{carroll2019spacetime}. By varying this action with respect to $\psi(x^\mu)$, one can derive the wave equations $\psi(x^\mu)^{;\alpha}_{~~;\alpha}=0$, which, to the first order of $\epsilon$, can be written as \begin{equation}
	[\eta ^{\alpha\beta}-\epsilon\theta(t) h^{\alpha\beta}(x^\mu)]\psi_{,\alpha\beta }(x^\mu)=0,
\end{equation}
where $\theta(t)$ is the Heaviside function and is used to restrict the interaction time to $t>0$. In the perturbation theory, we can write the scalar field as $\psi(x^\mu)=\psi^{(0)}(x^\mu)+\epsilon \psi^{(1)}(x^\mu)$, where $\psi^{(0)}(x^\mu)$ is the unperturbed field satisfying $\psi^{(0)}(x^\mu)^{,\alpha}_{~~,\alpha}=0$, $\psi^{(1)}(x^\mu)$ is the perturbation part, and the perturbation parameter $\epsilon$ indicates that $\psi^{(1)}(x^\mu)$ are corrections that are proportional to the perturbation metric field $h_{\alpha\beta}(x^\mu)$ \cite{sakurai2020modern,exirifard2021towards}. To the first order of $\epsilon$, the equations of motion (EOMs) for $\psi^{(1)}(x^\mu)$ is given by\begin{equation}
	\eta^{\alpha\beta}\psi^{(1)}(x^\mu)_{,\alpha\beta}=D_{\alpha\beta}h^{\alpha\beta}(x^\mu)=\theta(t)\psi^{(0)}(x^\mu)_{,\alpha\beta}h^{\alpha\beta}(x^\mu),\label{eq.EOMofPerturbationField}
\end{equation}
where $D_{\alpha\beta}=\theta(t)\psi^{(0)}(x^\mu)_{,\alpha\beta}$ is referred to as the antenna patter or the detector response in GW detectors. The unperturbed scalar field $\psi^{(0)}$ can be decomposed by Bessel modes as \cite{khonina2020bessel} 
\begin{align}\psi^{(0)}=&\sum_{l=-\infty}^\infty \int_0^\infty d p_\perp \int_{-\infty}^\infty dp_3 \Big (g_{p_3p_\perp l}(x^\alpha)a_{p_3p_\perp l}\nonumber \\&+g^*_{p_3p_\perp l}(x^\alpha)a_{p_3p_\perp l}^* \Big ),\label{eq.fieldInFlat}\end{align} where $a_{p_3p_\perp l}$ are coefficients. The Bessel modes, in the cylindrical coordinates $(t,\rho,\phi,z)$, are given by \begin{equation}
	g_{k_3 k_\perp l}(x^\alpha)=C_{k_3 k_\perp l}J_l(k_\perp \rho)e^{i(-E t+k_3 z+l\phi)},\label{eq.BesselMode}
\end{equation} where $C_{k_3 k_\perp l}=[k_\perp/(8 \pi^2 E)]^{1/2}$ is the normalization factor, $k_3 $ is the wavevector along $z$ axis, $k_\perp$ is the radial wavevector, $l$ is the OAM (also known as the topological charge), $J_l(x)$ is the $l$th-order Bessel function, and $E=(k^2_3+k^2_\perp)^{1/2}$. Here, we only examine solutions with positive norms ($E>0$---i.e., waves propagate along positive $z$ axis).

By comparing Eq. \eqref{eq.waveEquation} and Eq. \eqref{eq.EOMofPerturbationField}, one can find that the perturbation field can be solved by using the Green's function as \begin{align}
	&\psi^{(1)}(x^\mu)\nonumber \\=&\int d^4 x^{\mu'}\mathcal G(x^\mu,x^{\mu'}) \theta(t')\psi^{(0)}(x^{\mu'})_{,\alpha'\beta'}h^{\alpha'\beta'}(x^{\mu'})\nonumber \\
	=&\int d^4 x^{\mu'}\mathcal G(x^\mu,x^{\mu'})\theta(t') \sum_{l=-\infty}^\infty \int_0^\infty d p_\perp \int_{-\infty}^\infty dp_3 \nonumber \\ &\times \Big (g_{p_3p_\perp l}(x^{\mu'})_{,\alpha'\beta'}a_{p_3p_\perp l}+g^*_{p_3p_\perp l}(x^{\mu'})_{,\alpha'\beta'}a_{p_3p_\perp l}^* \Big )\nonumber \\ &\times h^{\alpha'\beta'}(x^{\mu'})\nonumber \\
	=&\sum_{l=-\infty}^\infty \int_0^\infty d p_\perp \int_{-\infty}^\infty dp_3 \Big (\tilde g_{p_3p_\perp l}(x^{\mu})a_{p_3p_\perp l}\nonumber \\ &+\tilde g^*_{p_3p_\perp l}(x^{\mu})a_{p_3p_\perp l}^* \Big ),
\end{align}
where the perturbation mode function $\tilde g_{k_3k_\perp l}(x^\alpha)$ is given by \begin{align}
	&\tilde g_{k_3k_\perp l}(x^\mu)\nonumber \\ =&\int d^4 x^{\mu'}\mathcal G(x^\mu,x^{\mu'})\theta(t')   
	h^{\alpha'\beta'}(x^{\mu'}) g_{k_3k_\perp l}(x^{\mu'})_{,\alpha'\beta'}\nonumber \\
	=&\int d^4 x^{\mu'}
	\frac{1}{8\pi^3}\sum_{m=-\infty} ^\infty\int_\infty^\infty dp_3 \int_0^\infty d p_\perp \int dp_0 \frac {p_\perp}{p^2_0-\omega^2} \nonumber \\ &\times e^{-ip_0(t-t')+ ip_3(z-z')+im(\phi-\phi')+i\phi_0}  J_m(p_\perp \rho) J_m(p_\perp \rho') \nonumber \\ &\times\theta(t-t')\theta(t')
	\epsilon^{\alpha'\beta'} e^{ik_g(-t'+z'\cos\theta)}\sum_{n=-\infty}^\infty i^n J_n(k_g \rho' \sin\theta) \nonumber \\ &\times e^{in\phi'} C_{k_3 k_\perp l}\left [ J_l(k_\perp \rho')e^{i(-E t'+k_3 z'+l\phi')} \right ]_{,\alpha'\beta'},\label{eq.tgBeforeIntegral}
\end{align}
$\omega=(p_3^2+p_\perp^2)^{1/2} $, $E=(k_3^2+k_\perp^2)^{1/2}$, and the Jacobi-Anger expansion, \begin{equation}
	\exp (ik_g x'\sin \theta)=\sum_{n=-\infty}^\infty i^n J_n(k_g\sin\theta \rho')\exp (in\phi'),
\end{equation}
is used \cite{lozier2003nist,olver2010nist}. By combining $\psi^{(0)}(x^{\mu'})$ and $\psi^{(1)}(x^\mu)$, one can find that, in the GWs, the field can be decomposed as
\begin{align}
	\psi_{\text{GW}}(x^\alpha)=&\sum_{j=-\infty}^\infty \int_0^\infty d k_\perp \int_{-\infty}^\infty dk_3 \Big (d_{k_3k_\perp j}(x^\alpha)b_{k_3k_\perp j}\nonumber \\ &+d^*_{k_3k_\perp j}(x^\alpha)b_{k_3k_\perp j}^* \Big ),\label{eq.fieldInGW}\end{align}
where $d_{k_3k_\perp j}(x^\mu)=g_{k_3k_\perp j}(x^\mu)+\tilde g_{k_3k_\perp j}(x^\mu)$ are mode functions in GWs, and the coefficients have been renamed as 
$b_{k_3k_\perp j}$ to be distinguished with the field in flat spacetime. Note the systematic perturbation symbol $\epsilon$ before $\tilde g_{k_3k_\perp l}(x^\alpha)$ has been set to 1, as the perturbation analysis has finished.

The integral in Eq. \eqref{eq.tgBeforeIntegral} with respect to $\rho'$ is rather complex. To provide an analytical solution, we impose the following assumptions:
\begin{equation}
	k_3\gg k_g,{\rm ~~and~~}k_\perp \gg k_g.
\end{equation}
The two assumptions suggest that the metric variation is negligible within the scales of $k_3^{-1}$ and $k_\perp^{-1}$. Under these assumptions, we have $\exp (ik_g x'\sin \theta)=\sum_{n=-\infty}^\infty i^n J_n(k_g\sin\theta \rho')\exp (in\phi')\approx 1$, and we can conclude that
\begin{align}
	&\tilde g_{k_3k_\perp l}(x^\alpha)\nonumber \\=&-\frac{C_{k_3k_\perp l}}{2 \omega_1}e^{i(k_g\cos\theta+k_3)z+i\phi_0}\bigg [ \frac 1 {\omega_1+k_g+E} \Big (e^{-i(k_g+E)t} \nonumber \\& -e^{i\omega_1t}\Big )+\frac 1 {\omega_1-k_g-E}  \Big (e^{-i(k_g+E)t}-e^{-i\omega_1t}\Big ) \bigg ] \Bigg [ b_2 \frac {k^2_\perp}{8}\nonumber \\ &\times J_{l+2}(k_\perp \rho) e^{i(l+2)\phi}+b_{-2} \frac {k^2_\perp}{8}J_{l-2}(k_\perp \rho) e^{i(l-2)\phi}\nonumber \\ &+b_1k_\perp k_3 J_{l+1}(k_\perp \rho) e^{i(l+1)\phi}+b_{-1}k_\perp k_3 J_{l-1}(k_\perp \rho) e^{i(l+1)\phi}\nonumber \\ & +b_0\left (k^2_3-\frac {k^2_\perp}{2} \right ) J_{l}(k_\perp \rho) e^{il\phi}\Bigg ],\nonumber \\
	\approx&\frac{C_{k_3k_\perp l}}{2 Ek_g} e^{ik_3z+i\phi_0}\Big (e^{-i(k_g+E)t}  -e^{-iE t}\Big )  \Bigg [ b_2 \frac {k^2_\perp}{8}J_{l+2}(k_\perp \rho)\nonumber \\ &\times e^{i(l+2)\phi}+b_{-2} \frac {k^2_\perp}{8}J_{l-2}(k_\perp \rho) e^{i(l-2)\phi}  +b_1k_\perp k_3\nonumber \\ &\times J_{l+1}(k_\perp \rho) e^{i(l+1)\phi}+b_{-1}k_\perp k_3 J_{l-1}(k_\perp \rho) e^{i(l+1)\phi}\nonumber \\ & +b_0\left (k^2_3-\frac {k^2_\perp}{2} \right ) J_{l}(k_\perp \rho) e^{il\phi}\Bigg ],
\end{align}
where \begin{align}
	b_0=&-A_+\sin^2\theta,\label{eq.b0}\\
	b_1=b_{-1}^*=&(A_\times+i A_+ \cos\theta)\sin\theta,\\
	b_2=b_{-2}^*=&A_+(3+\cos2\theta)-4iA_\times\cos\theta,\label{eq.b2}
\end{align}
$\omega_1=[(k_g\cos\theta+k_3)^2+k_\perp^2]^{1/2}\approx E$, the assumption that $k_3\gg k_g$ is used, and the term proportional to $(2E+k_g)^{-1}$ is ignored. Note that the information of GW strain and polarization is encoded in the coefficients $b_i$ with $i=-2$ to $2$. As stated before, $\tilde g_{k_3k_\perp l}(x^\alpha)$ should be derived again with $k_g$ and $\phi_0$ replaced by $-k_g$ and $-\phi_0$, respectively. Then, for \begin{equation}
h^{\rm TT}_{\alpha\beta}(x^\mu)= \epsilon_{\alpha\beta}\cos [k_g(-t+x\sin\theta +z\cos\theta )+\phi_0],
\end{equation}
$\tilde g_{k_3k_\perp l}(x^\alpha)$ is the average of the two cases---i.e., \begin{align}
	\tilde g_{k_3k_\perp l}(x^\alpha)\nonumber =&\frac{-i C_{k_3k_\perp l}}{2 Ek_g} e^{ik_3z-iE t}[\sin (k_g t-\phi_0)+\sin \phi_0]\nonumber \\ &\times\Bigg [ b_2 \frac {k^2_\perp}{8}J_{l+2}(k_\perp \rho) e^{i(l+2)\phi}+b_{-2} \frac {k^2_\perp}{8}\nonumber \\ &\times J_{l-2}(k_\perp \rho) e^{i(l-2)\phi} +b_1k_\perp k_3 J_{l+1}(k_\perp \rho)\nonumber \\ &\times e^{i(l+1)\phi}+b_{-1}k_\perp k_3 J_{l-1}(k_\perp \rho) e^{i(l-1)\phi}\nonumber \\ & +b_0\left (k^2_3-\frac {k^2_\perp}{2} \right ) J_{l}(k_\perp \rho) e^{il\phi}\Bigg ].\label{eq.tildeg}
\end{align}

\section{Evolution of quantum states in GWs}
According to the canonical quantization procedure in quantum field theory, in flat spacetime, the scalar field is quantized as \cite{birrell1984quantum,chakraborty2024particle}\begin{align}
	\psi_{\text{FL}}(x^\alpha)=&\sum_{l=-\infty}^\infty \int_0^\infty d p_\perp \int_{-\infty}^\infty dp_3 \Big (g_{p_3p_\perp l}(x^\alpha){\hat a}_{p_3p_\perp l}\nonumber \\ &+g^*_{p_3p_\perp l}(x^\alpha){\hat a}_{p_3p_\perp l}^\dagger \Big ),
\end{align}
while in the presence of GWs, the field is quantized as\begin{align}
	\psi_{\text{GW}}(x^\alpha)=&\sum_{j=-\infty}^\infty \int_0^\infty d k_\perp \int_{-\infty}^\infty dk_3 \Big (d_{k_3k_\perp j}(x^\alpha){\hat b}_{k_3k_\perp j}\nonumber \\ &+d^*_{k_3k_\perp j}(x^\alpha){\hat b}_{k_3k_\perp j}^\dagger \Big ),
\end{align}
where the coefficients $a_{p_3p_\perp l}$ and $b_{k_3k_\perp j}$ in Eqs. \eqref{eq.fieldInFlat} and \eqref{eq.fieldInGW} are promoted to operators ${\hat a}_{p_3p_\perp l}$ and ${\hat b}_{k_3k_\perp j}$, respectively. Then, the Bogoliubov transformation between ${\hat a}_{p_3p_\perp l}$ and ${\hat b}_{k_3k_\perp j}$ is given by \cite{chakraborty2024particle}\begin{align}
	{\hat b}_{k_3k_\perp j}=&\sum_{l=-\infty}^\infty \int_0^\infty d p_\perp \int_{-\infty}^\infty dp_3 \Big (\alpha^*_{k_3k_\perp j,p_3p_\perp l}{\hat a}_{p_3p_\perp l}\nonumber \\ &-\beta^*_{k_3k_\perp j,p_3p_\perp l}{\hat a}^\dagger_{p_3p_\perp l} \Big ),\label{BTransformForbOperator}
\end{align}
with $\alpha_{k_3k_\perp j,p_3p_\perp l}=\left <g_{p_3p_\perp l}(x^\alpha) ,d_{k_3k_\perp j}(x^\alpha)\right >$ and $\beta_{k_3k_\perp j,p_3p_\perp l}=-\left <g^*_{p_3p_\perp l}(x^\alpha) ,d_{k_3k_\perp j}(x^\alpha)\right >$ being the Bogoliubov coefficients. The coefficient $\beta_{k_3k_\perp j,p_3p_\perp l}$ is related to the particle creation in GWs \cite{chakraborty2024particle}. This is an extremely weak process, when compared to the quantum transition in our work. Therefore, we will not consider $\beta_{k_3k_\perp j,p_3p_\perp l}$ furthermore. The Klein-Gordon inner product $\left < \cdot,\cdot\right >$ is defined as \cite{carroll2019spacetime}
\begin{equation}
	\left <g_{I},d_{J}\right >=i\int_\Sigma \big [g_{I}^*\partial_\mu d_{J} - d_{J} \partial_\mu g_{I}^*\big ]n^\mu \sqrt \gamma d^3x, \label{eq.innerProduct}
\end{equation}
where $g_{I}=g_{I}(x^\mu)$, $d_{J}=d_{J}(x^\mu)$, $I$ and $J$ represent a set of possible indices, $\Sigma$ is a spacelike hypersurface on which the integration is carried out, $\gamma_{\mu\nu}$ is the induced 3-metric on this hypersurface, $\gamma=\det( \gamma_{\mu\nu})$, and $n^\mu$ is the unit normal vector. In our case, to the first order of the metric perturbation $h_{\mu\nu}$, we have $n^\mu=(1,0,0,0)$ and $\gamma= 1$, so the inner product is calculated in Minkowski spacetime, and the GW information is encoded in the perturbation field mode function $\tilde g_{k_3k_\perp l}(x^\alpha)$. If both $g_I$ and $d_J$ satisfy the wave equation in Minkowski spacetime, it can be shown that the current $J_\mu= g_{I}^*\partial_\mu d_{J} - d_{J} \partial_\mu g_{I}^*$ is conserved---i.e., $J^\mu_{~~,\mu}=0$, so the inner product, Eq. \eqref{eq.innerProduct}, does not depend on the choice of $\Sigma$, provided that $J_\mu$ vanishes at spatial infinity \cite{takagi1986vacuum}. Therefore, we could choose $\Sigma$ to be the hypersurface of $t=0$. However, when calculating the Bogoliubov coefficients, the mode function $d_{k_3k_\perp j}(x^\alpha)$ does not satisfy the wave equation in Minkowski spacetime. To be concrete, the perturbation part $\tilde g_{p_3p_\perp l}(x^\alpha)$ depends on a time-dependent source term, so the Bogoliubov coefficients will depend on the time coordinate, as well. By calculating the inner product, one may find that \begin{align}
	&\alpha_{k_3k_\perp j,p_3p_\perp l}\nonumber \\=&\delta(k_3-p_3)\delta(k_\perp -p_\perp)\delta(j-l)+\big (-2iE[\sin(k_g t-\phi_0)\nonumber \\ &+\sin \phi_0]+k_g\cos(k_g t-\phi_0) \big )/(4E^2k_g)
 \delta(k_3-p_3) \nonumber \\ &\times\delta(k_\perp-p_\perp)\Bigg [ \frac {b_2 k^2_\perp}{8}\delta(j+2-l)+ \frac {b_{-2}k^2_\perp}{8}\delta(j-2-l)\nonumber \\ &+b_1k_\perp k_3 \delta(j+1-l)+b_{-1}k_\perp k_3 \delta(j-1-l)\nonumber \\ &+b_0\left (k^2_3-\frac {k^2_\perp}{2} \right )\delta(j-l)\Bigg ].\label{eq.alpha}
\end{align}

When deriving $\tilde g_{p_3p_\perp l}(x^\alpha)$ as given by Eq. \eqref{eq.tildeg}, we have ignored the variation of $k_3$ and $k_\perp$, since $k_3\gg k_g $ and $k_\perp\gg k_g$ are assumed. These assumptions further lead to the common factors $\delta(k_3-p_3)\delta(k_\perp-p_\perp)$ in Eq. \eqref{eq.alpha}. When $\alpha_{k_3k_\perp j,p_3p_\perp l}$ is substituted into Eq. \eqref{BTransformForbOperator}, the integrals with respect to $p_3$ and $p_\perp$ will let $p_3\rightarrow k_3$ and $p_\perp \rightarrow k_\perp$. Therefore, Eq. \eqref{BTransformForbOperator} indicates that the annihilation operators in GWs, $\hat b_{k_3k_\perp l}$, are only connected to those operators in flat spacetime with the same wavevector $k_3$ and $k_\perp$---i.e., $\hat a_{k_3k_\perp j}$, where $j$ may be different from $l$. As a result, we will focus on the OAM space and, for simplicity, we will write 
\begin{align}
		{\hat b}_{k_3k_\perp j}^\dagger=&\sum_{l=-\infty}^\infty \int_0^\infty d p_\perp \int_{-\infty}^\infty dp_3 \alpha_{k_3k_\perp j,p_3p_\perp l}{\hat a}_{p_3p_\perp l}^\dagger\nonumber \\ 
	=&\sum_{l=-\infty}^\infty \alpha_{j,l}{\hat a}_{k_3k_\perp l}^\dagger,
\end{align}
with\begin{equation}
\alpha_{j,l}=\int_0^\infty d p_\perp \int_{-\infty}^\infty dp_3 \alpha_{k_3k_\perp j,p_3p_\perp l}.\label{eq.alphajl}	
\end{equation}
Assume that at $t=0$, a photon with a wave vector $(k_3, k_\perp)$ exists in the OAM state $\ket{l}$. As the photon begins to interact with the GW at $t=0$, it will evolve according to the mode functions $d_{k_3k_\perp j}(x^\alpha)$, so its state can be written as ${\hat b}^\dagger_{k_3,k_\perp, l} \ket 0$. Suppose the light is not reflected by mirrors. At a later time $t$, we will measure its state by using the mode functions of Bessel beams $g_{k_3k_\perp j}(x^\alpha)$. Consequently, the probability amplitude for the photon to be in a state $\ket{l'}$ is given by \begin{equation}
	\braket{l'}{l}=\bra 0 {\hat a}_{k_3,k_\perp, l'}{\hat b}^\dagger_{k_3,k_\perp,l}\ket 0=\alpha_{l,l'}.
\end{equation}
Therefore, the transition probability is given by $P_{l'}=\left |\alpha_{l,l'} \right |^2$. By using Eq. \eqref{eq.alphajl} to calculate $P_{l'}$, one may notice that there are five different Kronecker delta functions, $\delta(j-m)$, with $m=l-2,~l-1,~l,~l+1,~{\rm and}~l+2$. Therefore, $P_{l'}$ is nonvanishing only when $l'=l-2,~\cdots,~ l+2$. This indicates that aside from the original OAM state $\ket{l}$, the photon may occupy one of four additional states $\ket{l\pm 2}$ and $\ket{l\pm 1}$, each with a probability of
\begin{align}
	P_{l\pm 2}=&f(t) \frac{k^4_\perp}{256 E^2 k^2_g}  [A^2_+(3+\cos 2\theta)^2+16 A^2_\times \cos^2\theta ], \label{eq.p11}\\
	P_{l\pm 1}=&f(t)\frac{k^2_\perp k^2_3}{4 E^2 k^2_g} (A^2_\times+A^2_+ \cos^2\theta)\sin^2\theta ,\label{eq.Probability1}
\end{align}
where $f(t)=[\sin \phi_0 -\sin (\phi_0-k_g t)]^2$. These probabilities are proportional to $A^2_+$ or $A^2_\times$---i.e., they are of the second order of the perturbative parameter $\epsilon$. When GWs are absent---i.e., $A_+=A_\times=0$, these probabilities, \eqref{eq.p11} and \eqref{eq.Probability1}, are zero. Certain features can be inferred from the aforementioned transition probabilities. First of all, aside from the original OAM state, the photon can transition to four more OAM states, as stated before. Second, although the unperturbed modes $g_{k_3k_\perp l}(x^\alpha)$ have been normalized, the new mode functions $d_{k_3k_\perp j}(x^\alpha)= g_{k_3k_\perp j}(x^\alpha) +\tilde g_{k_3k_\perp j}(x^\alpha)$ may not, and a new normalization factor may be needed. However, the perturbation solutions $\tilde g_{k_3k_\perp j}(x^\alpha)$ are proportional to the GW amplitudes $A_+$ or $A_\times$, so they are very small. If one tries to calculate the new normalization factor, it will turn out to be very close to 1, so we do not have to normalize the mode functions $d_{k_3k_\perp j}(x^\alpha)$. As a result, the probability that photons remain in the original OAM state is simply approximated by $P_l=1-P_{l+2}-P_{l+1}-P_{l-1}-P_{l-2}$. Third, all probabilities oscillate with the frequency of $k_g$. For low-frequency GWs, $f(t)\approx \cos^2 \phi_0 k^2_g t^2$; conversely, for high-frequency GWs, $f(t)$ attains a maximal value at $t=(\phi_0-\pi/2-k\pi)/k_g$, where $k$ is an integer, or a minimal value at $t=2k\pi/k_g$ and $t=(2k\pi-\pi+2\phi_0)/k_g$. This fact may be utilized to enhance or attenuate the detection of a specific frequency of GWs. Also, in most instances, $k_3 \gg k_\perp, k_g$, resulting in $P_{l\pm 1}\propto k^2_\perp $, and $P_{l\pm 2}\propto k^4_\perp /k^2_3$. Hence, $P_{l\pm 1} \gg P_{l\pm 2}$---i.e., the greater the OAM that photons gain or lose, the more challenging the process becomes. For a typical scenario, e.g., at a propagation distance of $L=1\times 10^7$ m, with photon wavelength $\lambda=700$ nm, $k_\perp=1\times 10^6~{\rm m}^{-1}$, $A_+=1\times 10^{-21}$, $A_\times=1\times 10^{-21}$, GW frequency $f_g=1~{\rm Hz}$, $\phi_0=0$, and $\theta=\pi/3$, the probabilities are $P_{l\pm1}\sim 10^{-17}$ and $P_{l\pm2}\sim 10^{-20}$. The probabilities can be enhanced by increasing the photon radial wave vector $k_\perp$ and the propagation distance $L$. Also, a GW with stronger amplitudes $A_{+,\times}$ and smaller frequency $f_g$ can significantly raise the probabilities. However, when $k_g L$ is large enough, the phase variation of GWs can not be ignored, and photons will pick up different phase factor after interacting with GWs. Some photons may interfere with others destructively, resulting a lower probabilities, as indicated by the function $f(t)$ in Eqs. \eqref{eq.p11} and \eqref{eq.Probability1}.

At first sight, $P_{l+ 2}=P_{l- 2}$, and $P_{l+ 1}=P_{l- 1}$, suggesting that the quantized OAM is exchanged among photons, with GWs merely augmenting this process. The polarization of GWs appears to remain unaltered. However, this is merely a superficial outcome resulting from our inappropriate choice of GW polarization basis. To elucidate the interaction between photon OAM and GWs, we must use the circular polarization basis. When we choose to use the circular polarization basis \cite{misner1973gravitation}, wherein \begin{align}
	A_+&=\frac 1 {\sqrt 2}(A_R+A_L),\\
	A_\times&=\frac i {\sqrt 2}(A_R-A_L),
\end{align}
and $A_R$ and $A_L$ represent the right-circular and left-circular component, respectively. When $A_+$ and $A_\times$ are substituted into Eqs. (\ref{eq.b0}-\ref{eq.b2}), one may have\begin{align}
	b_0=&\frac i {\sqrt 2}(A_R+A_L)\sin^2\theta,\\
	b_{\pm 1}=&\frac {\pm i}{\sqrt 2}[A_R(\cos\theta\pm 1)+A_L(\cos\theta \mp 1)]\sin\theta,\\
	b_{\pm 2}=&\frac 1 {\sqrt 2} [A_R(3+\cos 2\theta \pm 4 \cos \theta)\nonumber\\&+A_L(3+\cos 2\theta \mp 4\cos\theta)].
\end{align}
Further, the nonvanishing $\alpha_{j,l}$ are given by\begin{align}
	\alpha_{l,l}=&\frac {-1}{\sqrt 2}(A_R+A_L)\sin^2\theta\left ( k^2_3-\frac{k^2_\perp}{2}\right ),\\
	\alpha_{l\pm1,l}=&\frac {\pm i}{\sqrt 2}[A_R(\cos\theta\pm 1)+A_L(\cos\theta \mp 1)]\sin\theta k_3 k_\perp,\\
	\alpha_{l\pm2,l}=&\frac 1 {8 \sqrt 2} [A_R(3+\cos 2\theta \pm 4 \cos \theta)\nonumber \\&+A_L(3+\cos 2\theta \mp 4\cos\theta)]k^2_\perp.
\end{align}
Lastly, one can derive the transition probabilities as\begin{align}
	P_{l\pm  1}=&f(t)\frac{k^2_\perp k^2_3}{8 E^2 k^2_g} \big [A_R(\cos \theta \pm 1)+A_L(\cos \theta\mp 1)\big ]^2\sin^2\theta ,\label{eq.Probability2}\\
	P_{l\pm 2}=&f(t) \frac{k^4_\perp}{512 E^2 k^2_g}\big [A_R(3+\cos 2\theta\pm 4\cos \theta)\nonumber \\& +A_L(3+\cos 2\theta\mp 4\cos \theta)\big ]^2.\label{eq.ProbabilityCir1}
\end{align}
As in the linear polarization basis, these probabilities are proportional to the second order of the perturbation metric field, so when GWs are absent, they will vanish. In the case of parallel propagation with $\theta=0$, the nonvanishing probabilities are $P_{l+ 2}\propto A^2_R$ and $P_{l- 2}\propto A^2_L$. It is evident that, as the linearized gravitational field is a massless spin-two field \cite{misner1973gravitation}, when the GW exhibits a right-circular polarization, whose time-average spin angular momentum (AM) is 2, it cannot acquire additional AM from the photons. Consequently, the transfer of a photon from an OAM state $l$ to $l-2$ is prohibited, as this would necessitate the photon releasing its OAM to an external object, but direct interactions between photons are not accounted for in this model and the GWs cannot absorb additional AM. Meanwhile, the transition from $l$ to $l+2$ is permitted. The increase of photon OAM originates not from other photons, but from GWs. Similar rules may be applied for GWs with left-circular polarization and transition from an OAM state $l$ to $l-2$. Therefore, the probabilities associated with linear polarization GWs, as expressed in Eqs. \eqref{eq.p11} and \eqref{eq.Probability1}, indicate that at a given moment, a photon absorbs AM from GWs with a certain probability, while another photon emits the same AM back to the GWs with the same probability. As a result, the time-average spin AM of GWs is not changed.

Last, we consider the case where the photons are bounced between mirrors for $N$ rounds. The length between the mirrors is $L$, so the total light-path length is $2NL$. For the $n$th forward path, the initial GW phase will be given by
\begin{equation}
	\tilde \phi_0=\phi_0-2(n-1)k_g L.
\end{equation}
Then, the time-dependent term in the $n$th forward path transition amplitudes $A_{n}^{\rm (F)}=\alpha_{l,l'}$---i.e.,\begin{equation}
	-2iE[\sin(k_g t-\phi_0)+\sin \phi_0]+k_g\cos(k_g t-\phi_0),
\end{equation}
is replaced by 
\begin{align}
	&-2iE[\sin(k_g t-\phi_0+2(n-1)k_g L)+\sin (\phi_0-2(n-1)\nonumber \\&\times k_g L)]+k_g\cos(k_g t-\phi_0+2(n-1)k_g L).
\end{align}
Similarly, for $n$th backward path, the initial phase is 
\begin{equation}
\tilde \phi_0=\phi_0-(2n-1)k_g L+k_g L\cos\theta,
\end{equation}
and the time-dependent term in the $n$th backward path transition amplitudes $A_{n}^{\rm (B)}$ is replaced by\begin{align}
	&\{ -2iE[\sin(k_g t-\phi_0+(2n-1)k_g L-k_g L\cos\theta)+\sin (\phi_0\nonumber \\&-(2n-1)k_g L+k_g L\cos\theta)]+k_g\cos(k_g t-\phi_0+(2n-1)\nonumber \\&
	\times k_g L-k_g L\cos\theta) \} e^{i\pi},
\end{align}
where the term $e^{i\pi}$ represents the phase change when photons are reflected by mirrors. Note that, in this case, other $\theta$'s in probability amplitudes is replaced by $\pi-\theta$. Then, the total transition amplitude to OAM state $\ket {l+j}$, $A_j$, is given by
\begin{equation}
	A_j=\sum_{n=1}^N A_{j}^{\rm (F)}+\sum_{n=1}^N A_{j}^{\rm (B)},
\end{equation}
where $A_{j}^{\rm (F)}$ is the amplitude for the $n$th forward path and $A_{j}^{\rm (B)}$ is that for the $n$th backward path. Take the transition to the OAM state $ \ket {l-1}$ as an example. The transition probability will be given by\begin{align}
	&P_{-1}=\left |A_{-1} \right |^2\nonumber \\=&\frac{k^2_3 k^2_\perp \sin^2\theta \sin^2 (N k_g L)}{E^2k^2_g \cos^2 (k_g L/2)}\Big [A^2_+ \cos^2 \theta \cos^2((\cos\theta-1)k_gL/2)\nonumber \\ &\times \cos^2((\cos\theta-2N)k_gL/2+\phi_0)+A^2_\times  \sin^2((\cos\theta-1)\nonumber \\ &\times k_gL/2)\sin^2((\cos\theta-2N)k_gL/2+\phi_0)  \Big ].\label{eq.probabilityWithReflection}
\end{align}

\section{Photonic single-arm GW detectors}
The above results suggest that interactions with GWs can alter the photon OAM. Therefore, we propose a photonic single-arm GW detector based on the quantum state transition of OAM induced by GWs. At the sending port, photons are prepared in the Bessel mode with OAM of 1 by passing a laser via a spatial light modulator (SLM). This can also be accomplished through the utilization of annular apertures, axicons, Fabry-Perot resonators, or alternative techniques \cite{allam2024conceptual}. These photons will be referred to as the original photons. When necessary, photons may be reflected multiple times between two mirrors that comprise the arm; alternatively, the SLM and photon detector constitute the arm. The photon detector may alternatively be located on the same side as the SLM. When GWs pass the arm, the original photons may interact with the GWs, potentially relinquishing their OAM with a probability of $P_{l- 1}$. We refer to these non-OAM photons as signal photons. At the detecting port, only these signal photons will be captured by the photon detector. It has been demonstrated that vortex modes can be used to improve the performance of ground-based GW detectors such as Virgo and LIGO \cite{PhysRevD.79.122002,PhysRevLett.105.231102}. Similarly, our detection technique can also be realized in Virgo or LIGO to enhance their detection ability.

\begin{figure} [tbhp]
	\centering
	\subfloat[][Phase for $l=0$]{%
		\includegraphics[height=0.29\linewidth]{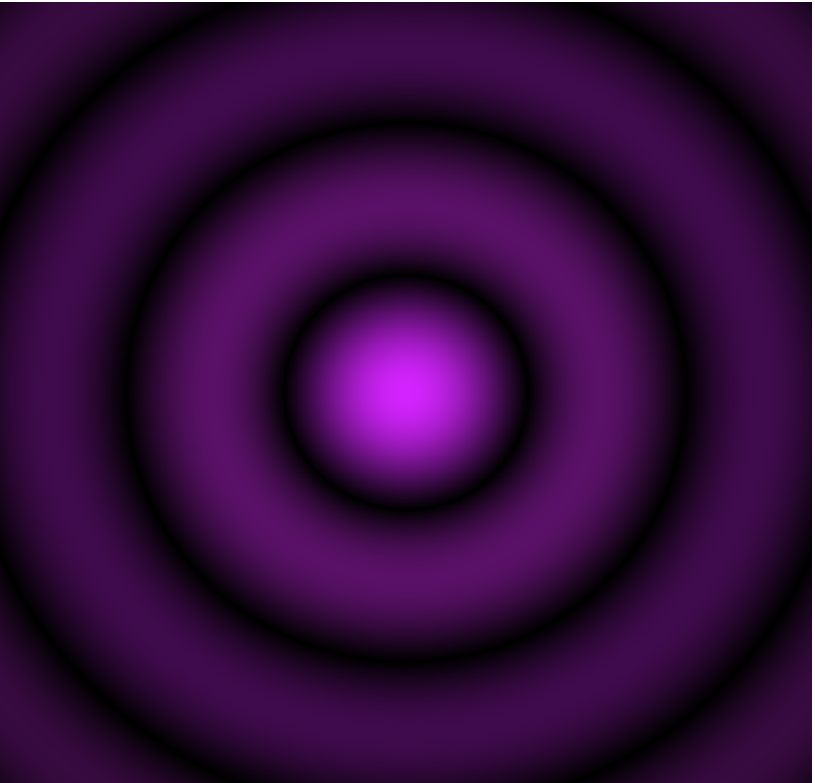}%
		\label{fig.0phase}
	}\hfill
	\subfloat[][Phase for $l=1$]{%
		\includegraphics[height=0.29\linewidth]{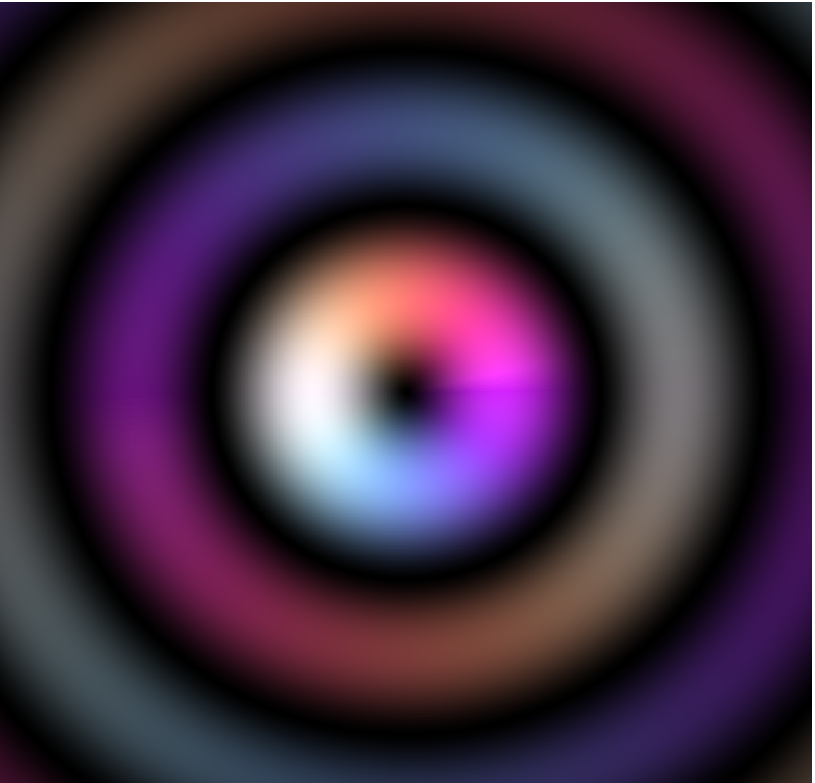}%
		\label{fig.1phase}
	}\hfill
	\subfloat[][Phase for $l=2$]{%
		\includegraphics[height=0.29\linewidth]{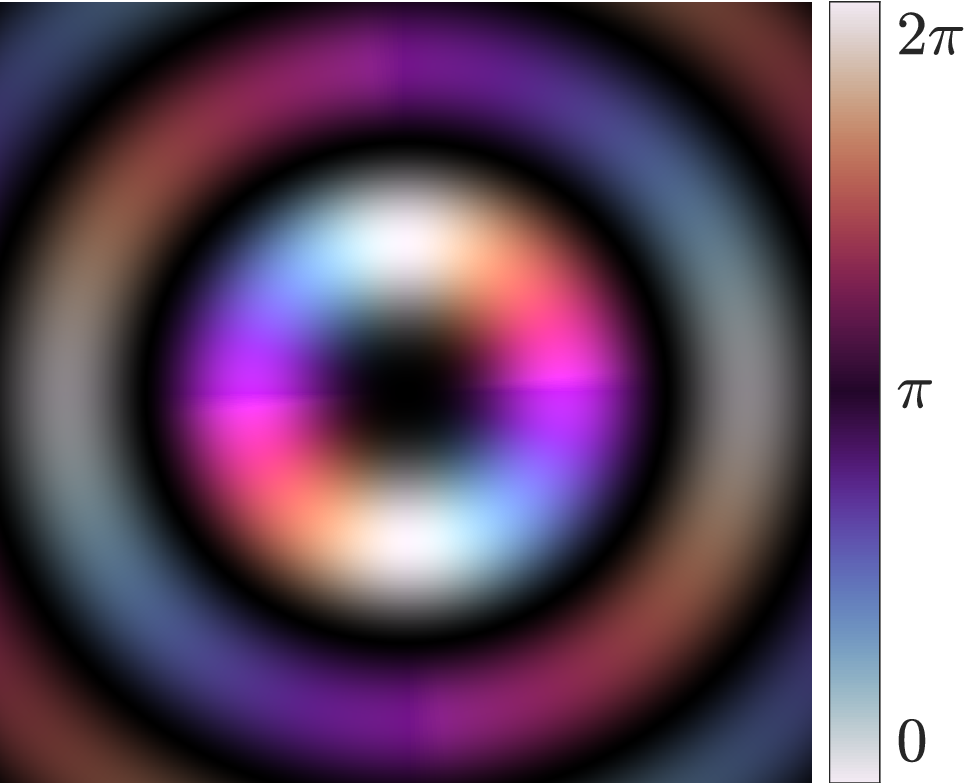}%
		\label{fig.2phase}
	}
	\caption{The phase for Bessel beams with $l=0$, 1, and 2, respectively. The brightness indicates field intensity.}
	\label{fig.OAMPattern}
\end{figure}

This design utilizes a significant property of vortex beams: When its OAM is nonzero, a central dark region appears along its propagation axis \cite{shen2019optical}, resulting from the phase singularity created by the helical wavefront, as depicted in Fig. \ref{fig.1phase} and \ref{fig.2phase}. Only the zeroth-order vortex beam can exhibit a central bright spot \cite{shen2019optical}, as shown in Fig. \ref{fig.0phase}, where the wavefront is a plane of uniform phase. The brightness in Fig. \ref{fig.OAMPattern} represents the field intensity, which are normalized by assigning the maximum intensity a value of one. Therefore, the original OAM photons will not be coupled to the photon detector; only photons with zero OAM will produce a central signal that can be captured by the photon detector, and this is why they are called the signal photons.

\begin{figure} [tbhp]
	\centering
	\includegraphics[width=1\linewidth]{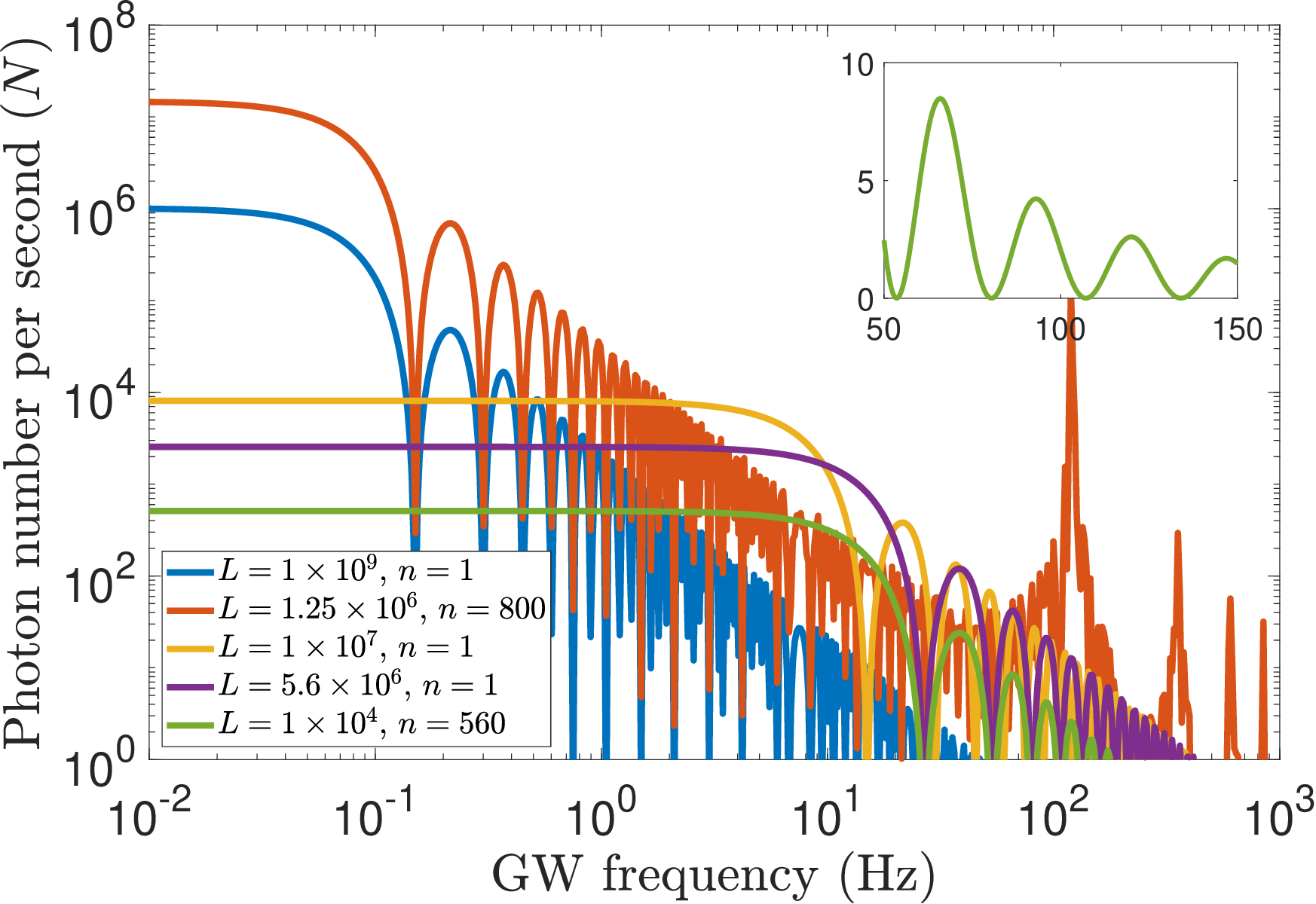}
	\caption{The rate of detected photon ($N$) for different GW frequencies ($10^{-2}$ to $10^3$ Hz) with a typical strain $A=10^{-21}$. The inset plot shows the behavior of $N$ between the GW frequency of 50 Hz and 150 Hz for the case where $L=1\times 10^4$ m and $n=560$, that can be achieved in a LIGO-like single-arm detector.}
	\label{fig.photonNumVFrequency}
\end{figure}

One may wonder about its ability to detect GWs. Here, the detection ability is quantified by the rate of detected photons ($N$), which is given by $N=P_{l- 1} P_{\rm rev}/\omega$, where $P_{\rm rev}$ is the received light power, and $\omega$ is the frequency of photons. For theoretical analysis, the photons are prepared in the Bessel beam modes. It is known that the Bessel beam is non-diffractive, but it can not be generated experimentally; instead, it may be replaced by Bessel-Gaussian beams or other types of modes. These beams will diverge as they propagate in space. At a distance $z$, it has a beam radius of $w(z)=w_0 [1+(z\lambda/(\pi w^2_0))^2]^{1/2}$ \cite{vallone2015properties}, where $w_0$ is the waist radius and $\lambda$ is the light wavelength. Assuming that the aperture at the detecting port has the size of $w_0$, the received light power will be given by $P_{\rm rev}=P_{\rm sed} w^2_0/w(z)^2$, with $P_{\rm sed}$ being the sending laser power. The interaction between photons and GWs is very weak, so a large waist radius $w_0$ is needed to ensure that enough signal photons will be received by the detector. Also, the light can be refocused at the mirrors (if they exist) to reduce the light divergence. Here, we choose $w_0=5$ m, which may be achieved by using a HF/DF chemical laser \cite{horkovich1997recent}. Figure \ref{fig.photonNumVFrequency} illustrates $N$ as the GW frequency ranges from $10^{-2}$ Hz to $10^4$ Hz with various arm lengths $L$ and counts of photon in-arm trips $n$. The parameters selected include a photon wavelength of $\lambda=700$ nm, a radial wavevector of $k_\perp=1\times 10^6~{\rm m}^{-1}$, a plus GW strain of $A_+=1\times 10^{-21}$, a cross GW strain of $A_\times=1\times 10^{-21}$, a laser power of $P_{\rm sed}=100$ W, an initial GW phase of $\phi_0=0$, and a GW direction of $\theta=\pi/3$. As is shown, the detector demonstrates exceptional performance, maintaining relatively steady $N$ within a specific frequency range, referred to as the steady frequency range. This occurs because of the ability of vortex photons to constantly make transitions into zeroth-order states with the same phase in the steady frequency range as they traverse the arm. As the total light path ($nL$) is increased, the steady frequency range diminishes, while $N$ significantly rises. 

In the case where $L=1\times 10^4$ m and $n=560$ (achievable in ground-base detectors), the GW frequencies can be categorized into three parts. The low-frequency part ($<1$ Hz) and the mid-frequency part ($1\sim10$ Hz) align with the steady frequency range, exhibiting a steady $N$ (about $ 500$ signal photons per second). It opens a potential window to detect GWs in the frequencies between 0.1 Hz and 10 Hz, which are difficult to identify in other contemporary interferometric detectors \cite{ni2018gravitational}. Hence, this detector may probe GWs from rich sources, including intermediate black hole binary coalescences, inspiral phases of stellar-mass coalescences, and compact binaries falling into intermediate black holes. In the high-frequency region ($>10$ Hz), $N$ is strongly affected by the GW frequency. At certain frequencies, the detector can identify some signal photons; however, in adjacent regions, the signal photons significantly decrease, leading to a selection rule for GW frequencies. For example, the detection rate peaks at 65.9, 93.0, 120.0, and 146.9 Hz, as is shown in the inset plot in Fig. \ref{fig.photonNumVFrequency}. This occurs because, within the high-frequency GWs, the original photons experience varying GW phases during their propagation in the arm, resulting in their transition to signal photons with different phases. The signal photons will interfere with themselves either constructively or destructively based on their relative phases, resulting in the function $f(t)$. Therefore, $N$ exhibits significant sensitivity to the GW frequency when other detector parameters are fixed. For space-based detectors, a much greater arm length can be realized. For example, in the case where $L=1\times 10^7$ m and $n=1$, the detector can provide much higher $N$ (about $ 10^4$ signal photons per second) in the low-frequency part ($<1$ Hz), compared to the ground-based case. The detector can also probe GWs in the mid-frequency range ($1\sim10$ Hz), with more than $800$ signal photons per second. In the high-frequency region ($>10$ Hz), its performance also strongly depends on the GW frequency.

It is worthy of noting that this detector also depends on the interference of photons, similar to other contemporary interferometric GW detectors \cite{pitkin2011gravitational}. While other interferometric detectors depend on the phase difference caused by the light path discrepancy in two arms, our detector design requires only a single arm, wherein the signal photons interfere with themselves. When these photons possess identical phase factors, they will interfere constructively, amplifying the signal in the photon detector. However, when the signal photons possess different phase factors, particularly in the high-frequency range of GWs, they may interfere with themselves destructively, resulting in the decay of $N$ in Fig. \ref{fig.photonNumVFrequency}. This fact offers a significant advantage for our detector, as it is insensitive to the arm length $L$, hence potentially mitigating some background noise, e.g., seismic noise \cite{pitkin2011gravitational}. Moreover, by examining the probability $P_{l- 1}$, one can find that $N$ is proportional to $\left |A \right |^2$, with $A$ being the GW amplitude. Hence, this detector is more advantageous for determining the distance of the source compared to current interferometer detectors.

\section{Conclusion}

We have examined the propagation of vortex beams in GWs by using the Green's function method and perturbation theory. The light beams are represented by a massless scalar field and are expanded with modes of Bessel beams. The perturbation light field is derived under certain reasonable assumptions. The Bogoliubov transformation is employed to elucidate the evolution of quantum OAM states in GWs by connecting the quantization procedures in flat spacetime and in GWs. This yields transition probabilities for four distinct final OAM states.

Considering the interaction between GWs and photon OAM and the observation that only the zeroth-order vortex beam can exhibit a central bright spot, we suggest a photonic single-arm GW detector. The detection ability is studied by the rate of detected photons, $N$. The detector demonstrates exceptional performance in the steady frequency range. For the case where $L=1\times 10^4$ m and $n=560$ (for ground-based detectors), or $L=1\times 10^7$ m and $n=1$ (for space-based detectors), the detector provides a high and steady $N$ in the low-frequency range ($<1$ Hz), opens a potential window to detect GWs in the mid-frequency range ($1\sim10$ Hz), that is lacking in other contemporary GW detectors, and leads to a selection rule for GW frequencies in the high-frequency range ($>10$ Hz).

This technique not only facilitates the extraction of GW information but also establishes a new pathway towards selecting GW signals. Furthermore, the observable effects under GWs we calculate, similar to the Hawking radiation and the Unruh effect, are also manifestations of the quantization of the light field in the context of curved spacetime. However, more work remains before this detector can be actualized. First, an OAM laser with a power of 100 W has recently been recorded in a laboratory \cite{ding2025100,chen2024100}. The rate of detected photons $N$ is directly proportional to the laser power. Hence, if OAM photons can be generated using much higher laser power, the ability of our detector will be considerably enhanced. Second, the vortex beam has been extensively used across various fields, including optical tweezers, quantum communication, quantum computing, and quantum cryptography. However, the phase structure of a vortex beam is susceptible to the environment, e.g., atmospheric turbulence \cite{roux2011infinitesimal,roux2015entanglement,paterson2005atmospheric,roux2013decay,rodenburg2012influence}, so the vortex light transmission distance remains inside the confines of a city \cite{peplow2014twisted,krenn2014communication}. In space, the vortex light can also be affected by gravitational fluctuations \cite{PhysRevD.106.045023}. Fortunately, given the arm length examined in our study, these adverse effects from gravitational fluctuations are negligible. Still, it should be experimentally verified that the laser can convey OAM information at a distance of $10^7$ m. Third, there are some reports on generating high-purity OAM modes. In Ref. \cite{PhysRevLett.131.183801}, a high OAM mode purity up to 96.9\% is achieved by hybrid integration of a two-dimensional semiconductor $\rm{WSe_2}$ with a spin-orbit-coupled microring resonator. Ref. \cite{Zhang:15} proposes and designs a graded-index fiber for the OAM light, with which the OAM purity is higher than 99.9\% and the intrinsic crosstalk is suppressed to be lower than -30 dB. In Ref. \cite{Sontag:22}, a novel technique for generating beams of light carrying OAM that increases mode purity and decreases singularity splitting by orders of magnitude is presented, where adjacent OAM coefficients may be declined to near $10^{-10}$. Still, more studies are needed to suppress leakage at the level of $10^{-17}$ from the $l = 1$ mode to the $l = 0$ mode, via techniques such as optical fiber design \cite{Sontag:22}, fiber structure \cite{Liu:22}, antenna design \cite{huang2019generation}, metasurfaces \cite{Feng_2025}, etc. Fourth, while our detector is insensitive to arm length variations and can disregard certain background noise, such as seismic noise \cite{pitkin2011gravitational}, it remains susceptible to other forms of noise, including thermal and quantum noise. The main noise source comes from shot noise that arises from the quantized detection of the output light field on the photo detector. The shot noise is equal to the square root of the average photon number $\bar n$. Furthermore, when the energy of the light beam is given, $\bar n$ is only related to the photon frequency. Therefore, $\bar n$ does not depend on the OAM, so does the shot noise, as indicated in Ref. \cite{Kaviani:20}. Thus, the signal-to-noise ratio (SNR) due to the shot noise is given by \cite{pitkin2011gravitational}
\begin{equation}
	{\rm SNR}\propto\frac{\bar n}{\Delta n}=\sqrt{\bar n},
\end{equation}
where $\Delta n$ is the statistical fluctuation of $\bar n$. The shot noise can be suppressed by increasing the photon number in the cavity. In contemporary interferometric detectors, however, this increase will raise the radiation pressure noise on the test masses (mirrors) that will drive the mirror positions and limit the sensitivity \cite{pitkin2011gravitational}. As our detection technique is insensitive to the arm length, we expect that the SNR will not be dominated by the radiation pressure noise. In addition, thermal photons can exist in light fields, causing optical thermal noise. However, this noise is related to the quantum noise (mainly shot noise) by\begin{equation}
	n_{\rm th}\approx 2e^{-\hbar\omega/(k_BT)}\left (1+\frac{\hbar \Omega}{k_BT}\right )\Delta n,
\end{equation}
where $k_B$ is the Boltzmann constant, $T$ is the temperature, and $\Omega$ is the field frequency fluctuation around $\omega$. It is shown that this noise is negligible when compared to the shot noise \cite{PhysRevLett.130.241401}. As our current work focuses on the interaction theory between the OAM scalar light fields and the GWs, and the exploration of the theoretical possibility to detect GWs based on the quantum transition of OAM photons in GWs, more studies on, for instance, generating high-purity OAM photons and noise budget analysis, are needed before this detector can be realized.

\section*{Acknowledgments}
This work is supported by National Key R$\&$D Program of China (2020YFC2201400) and National Natural Science Foundation of China (12034016).

\bibliography{main.bbl}
\end{document}